\documentclass[twocolumn,prl,superscriptaddress]{revtex4}
\usepackage{graphicx}% Include figure files
\usepackage{rotate}

\usepackage{times}

\begin{document}

\title{Self-similar community structure in organisations}

\author{R. Guimer\`a}

\affiliation{Departament d'Enginyeria Qu\'{\i}mica, Universitat Rovira
i Virgili, 43007 Tarragona, Catalunya, Spain}

\affiliation{Department of Chemical Engineering, Northwestern
University, 60208 Evanston, IL, USA}

\author{L. Danon}

\affiliation{Departament de F\'{\i}sica Fonamental, Universitat de
Barcelona, 08028 Barcelona, Catalunya, Spain}

\affiliation{Departament d'Enginyeria Inform\`atica i Matem\`atiques,
Universitat Rovira i Virgili, 43007 Tarragona, Catalunya, Spain}

\author{A. D\'{\i}az-Guilera}

\affiliation{Departament de F\'{\i}sica Fonamental, Universitat de
Barcelona, 08028 Barcelona, Catalunya, Spain}

\affiliation{Departament d'Enginyeria Qu\'{\i}mica, Universitat Rovira
i Virgili, 43007 Tarragona, Catalunya, Spain}

\author{F. Giralt}

\affiliation{Departament d'Enginyeria Qu\'{\i}mica, Universitat Rovira
i Virgili, 43007 Tarragona, Catalunya, Spain}

\author{A. Arenas} 

\affiliation{Departament d'Enginyeria Inform\`atica i Matem\`atiques,
Universitat Rovira i Virgili, 43007 Tarragona, Catalunya, Spain}

\begin{abstract}
The formal chart of an organisation is designed to handle routine and
easily anticipated problems, but unexpected situations arise which
require the formation of new ties so that the corresponding extra
tasks can be properly accomplished. The characterisation of the
structure of such {\it informal networks} behind the formal chart is a
key element for successful management. We analyse the complex e-mail
network of a real organisation with about 1,700 employees and
determine its community structure. Our results reveal the emergence of
self-similar properties that suggest that some universal mechanism
could be the underlying driving force in the formation and evolution
of informal networks in organisations, as happens in other
self-organised complex systems.
\end{abstract}

\maketitle

%%%%%%%%%%%%%%%%%%%%%%%%%%%%%%%%%%%%%%%%%%%%%%%%%%%
Although the formal chart of an organisation is intended to prescribe
how employees interact, ties between individuals arise for personal,
political, and cultural reasons \cite{krackhardt93}. The understanding
of the formation and structure of such informal networks are key
elements for successful management
\cite{krackhardt93,mayo49,morgan97}. The traditional way of
investigating informal networks within organisations consists of
conducting surveys using employee questionnaires. However, employees
answers often contain subjective elements such as ``political''
motives and the worry about offending colleagues. Another significant
limitation of the questionnaire based analysis is that time and effort
costs make it prohibitively expensive to map the entire network even
for medium sized organisations. The rapid development of electronic
communications provides a powerful alternative to the traditional
analysis of informal networks. Indeed, the exchange of e-mails between
individuals in organisations reveals how people interact and allows
mapping the informal network in a non-intrusive, objective, and
quantitative way.

We surmise that the exchange of e-mails between individuals in
organisations reveals how people interact \cite{ebel??,adamic??} and
therefore provides a map of the real network structure behind the
formal chart. We analyse the complex e-mail network of a real
organisation with about 1,700 employees and determine its community
structure \cite{amaral00,albert02,dorogovtsev02,girvan02}. Our results
reveal the emergence of self-similar properties that suggest that some
universal mechanism could be the underlying driving force in the
formation and evolution of informal networks in organisations, as
happens in other self-organised complex systems \cite{banavar99}.

%\section{Preliminary analysis of the e-mail network}\label{build}
\begin{figure}[tb]
\centerline{\includegraphics*[width=0.7\columnwidth]{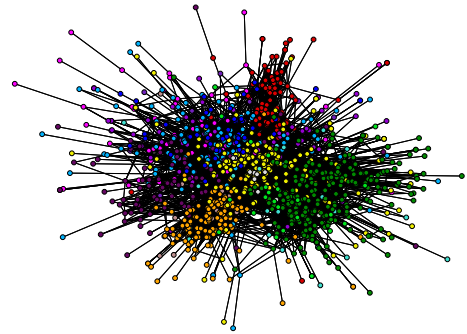}}
\caption{The e-mail network of URV. The network comprises
approximately 1700 users, including faculty, researchers, technicians,
managers, administrators, and graduate students. We consider e-mails
exchanged between university addresses during the first three months
of 2002. Each individual is represented by a node, with two
individuals (A and B) being connected if A has sent an e-mail to B and
B has also sent an e-mail to A. Bulk e-mails provide little or no
information about how individuals or teams collaborate. To minimise
their effect: (i) we eliminate e-mails that are sent to more than 50
different recipients and (ii) we disregard links that are
unidirectional, that is we consider only e-mails that represent a real
communication link, where e-mails flow in both directions. With these
two restrictions, the network is undirected and is formed by a main
component comprising 1133 nodes and many isolated nodes or pairs of
nodes. These little islands are not plotted to keep the figure as
simple as possible. The colour of each node identifies an individual's
affiliation to a specific centre within the university.}
\label{xarxa}
\end{figure}

Every time that an e-mail is sent, the addresses of the sender and the
receiver are routinely registered in a server. Therefore, an {\it
e-mail network} can be built regarding each e-mail address as a node
and linking two nodes if there is an e-mail communication between
them. In particular, we take as a case study the e-mail network of
University Rovira i Virgili (URV) in Tarragona, Spain, containing
around 1700 users (Fig. \ref{xarxa}). Bulk e-mails provide little or
no information about how individuals or teams collaborate and, once
they are removed, the connectivity distribution of the e-mail network
is exponential, $P(k)\propto\exp (-k/k^*)$ for $k\ge 2$, with $k^*=
9.2$. This result is in contrast with recent findings indicating that
some technology based social networks---such as rough e-mail networks
\cite{ebel??}, the Instant Messaging Network \cite{smith??} or the PGP
encryption network \cite{guardiola??}---show heavily skewed degree
distributions, but is consistent with the proposal of Amaral and
coworkers that the truncation of the scale-free behaviour in real
world networks is due to the existence of limitations or costs in the
establishment of connections \cite{amaral00}. Indeed, it seems
plausible that there are costs to maintaining active social
acquaintances and therefore active communications. However, it is
relatively easy to keep many {\it electronic} acquaintances {\it
open}, although most of them are probably inactive from a social point
of view.

\begin{figure}
\centerline{\includegraphics*[width=\columnwidth]{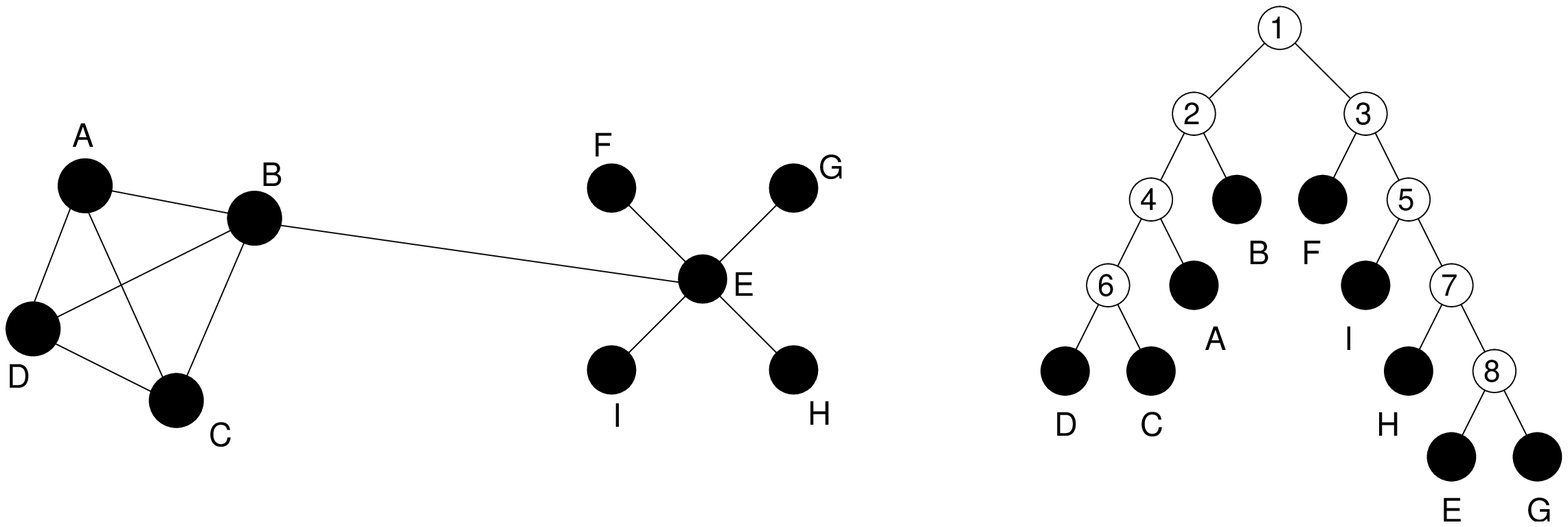}}
\centerline{{\bf a}\hspace{0.5\columnwidth}{\bf b}}
%\vspace*{0.5cm}
\caption{Community identification according to the GN algorithm. {\bf
a}, The betweenness of an edge is defined as the number of minimum
paths connecting pairs of nodes that go through that edge
\cite{wasserman94,newman01}. The GN algorithm is based on the idea
that the edges which connect highly clustered communities have a
higher edge betweenness---in this case, edge $BE$---and therefore
cutting these edges should separate communities. The algorithm
proceeds by identifying and removing the link with the highest
betweenness in the network. After every removal, the betweenness of
the edges is recalculated. This process is repeated until the `parent'
network splits, producing two separate `offspring' networks. The
offspring can be split further in the same way until they comprise of
only one individual. {\bf b}, In order to describe the entire
splitting process, we generate a binary tree, in which bifurcations
(white nodes) depict communities and leaves (black nodes) represent
individual addresses of the e-mail network. At the beginning of the
process, the network is a single entity, represented by node 1 in the
tree. After the removal of the edge $BE$, the network is split into
two subnetworks, 2 and 3, containing addresses A to D and E to I
respectively. The two offspring networks have no further internal
community structure. Consider first, subnetwork 2 containing nodes A
to D. When all the links are equivalent and have the same betweenness
as in the present case, one of them will be selected at random for
removal. It is straightforward to show that, iterating the link
removal procedure, nodes will be separated one by one and randomly by
the GN algorithm, generating a branch in the binary tree. As an
example, the figure represents a situation in which $B$ is separated
first, then $A$, and finally $D$ and $C$, but a different random
selection of links would lead to a different separation
order. Similarly, in subnetwork 3 nodes will be separated one by one
and at random, except for the fact that the most central node, $E$,
will always be separated last. In general, for large networks in which
the probability of having two links with the same betweenness is very
small, it will still be true that communities will appear as branches
in the community binary tree and that the tips of the branches will
correspond to the most central agents in the network.}
\label{algorithm}
\end{figure}
To understand the structure of the informal network of the
organisation, we are interested in determining how individuals
interact and form groups that, in turn, interact with each other
giving rise to higher order groups, that is, groups of groups. In
other words, we want to unravel the {\it community structure} of the
network. To do so we use the algorithm proposed recently by Girvan and
Newman (GN) \cite{girvan02} to identify communities in complex
networks (see Fig. \ref{algorithm}). The algorithm proceeds by
splitting the network recursively until single nodes are left. The
information about the community structure of the original network can
be deduced from the topology of the binary tree that represents this
splitting procedure and which leaves correspond to addresses of the
e-mail network (Fig. \ref{algorithm}b). The different communities of
the original network appear as branches in this tree, which are easily
identified by visual inspection.
%
%Moreover, the nodes that appear at the end of the branches are, in
%general, the most central nodes of their corresponding communities.
%
\begin{figure}[h!]
\centerline{\includegraphics*[width=0.90\columnwidth]{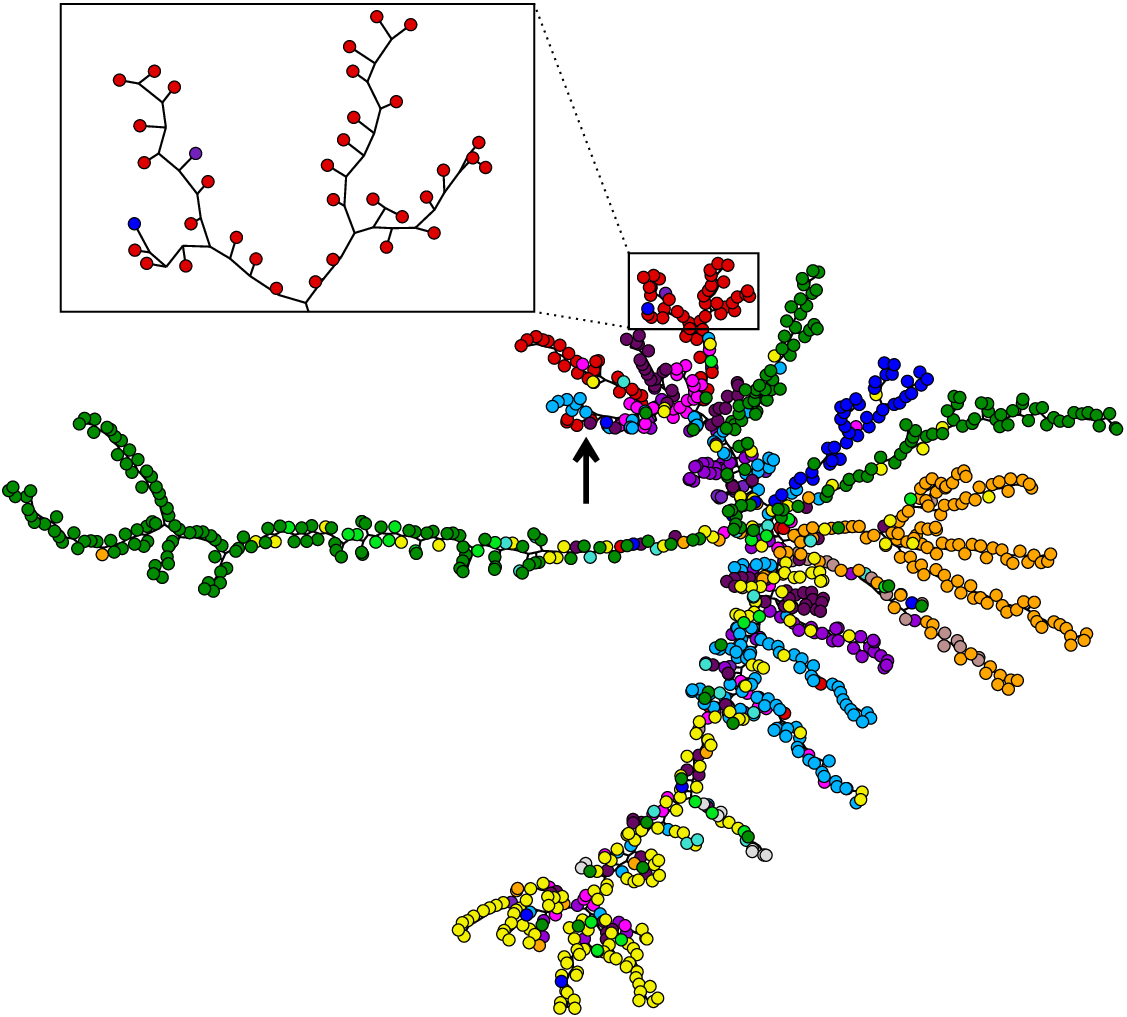}}
%\vspace{0.25cm}
\centerline{{\bf a}}
\vspace{.9cm}
\centerline{\includegraphics*[width=0.4\columnwidth]{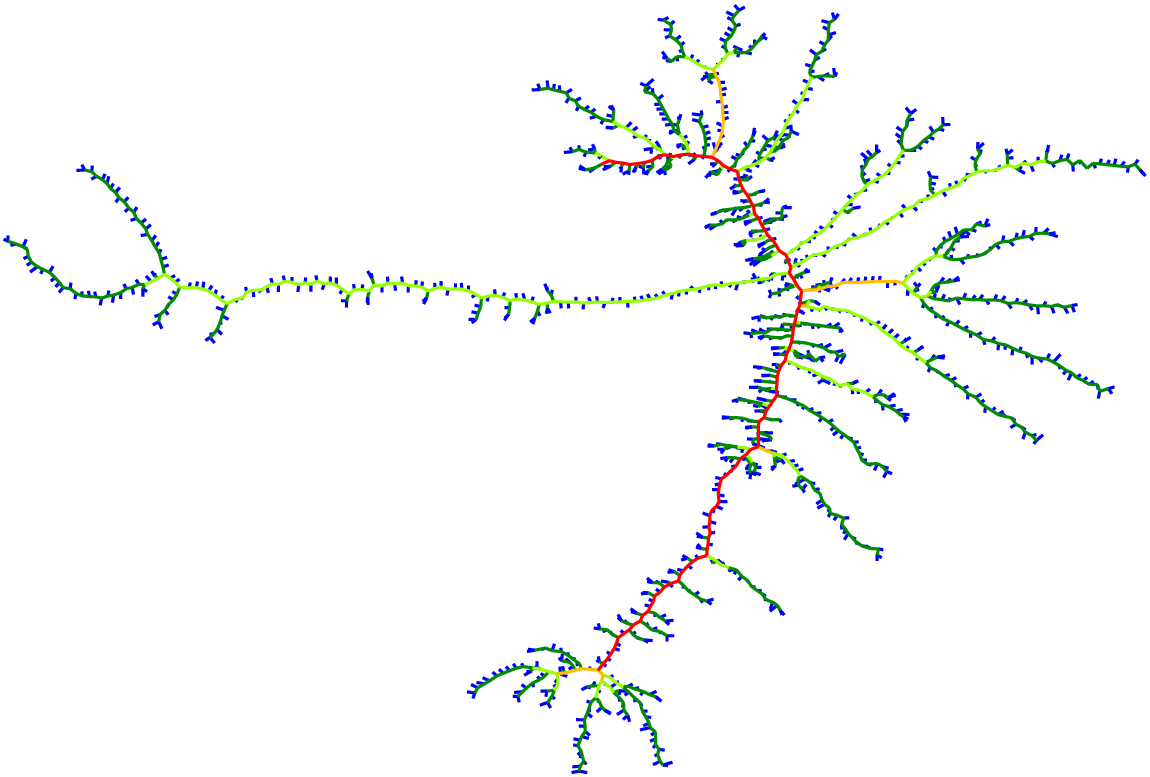}\quad\includegraphics*[width=0.40\columnwidth]{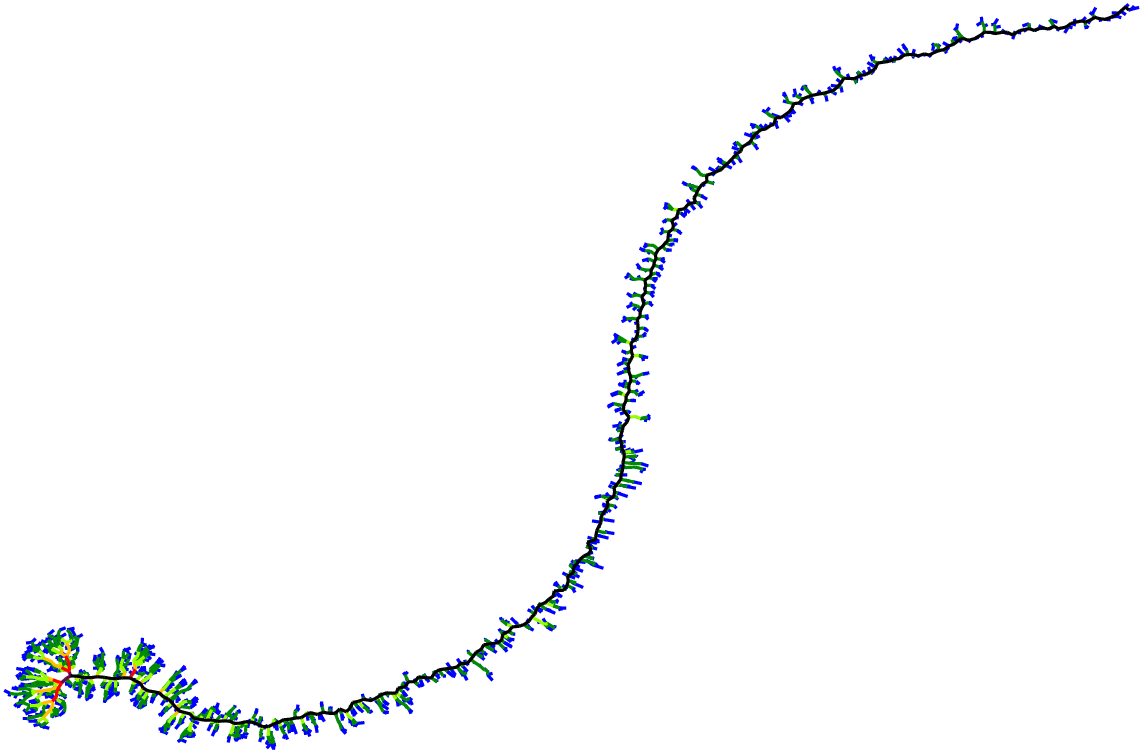}\quad\includegraphics*[width=0.05\columnwidth]{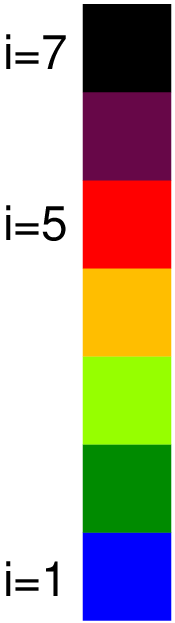}}
\centerline{{\bf b}\hspace{0.35\columnwidth}{\bf c}}
\caption{Communities in the e-mail network of URV. {\bf a}, Binary
tree showing the result of applying the GN algorithm to the e-mail
network of URV. The position indicated by the arrow represents the
root of the tree (equivalent to node 1 in figure \ref{algorithm}b) and
branches are depicted so that they can be clearly differentiated. In
particular, only the leaves of the tree, that correspond to e-mail
addresses, are plotted, as shown in the detail that is zoomed. The
colour of each of the leaves represents different centres within the
university (five small centres containing less than 10 individuals are
assigned the same colour). Nodes of the same colour (from the same
centre) tend to stick together in the same branch meaning that
individuals within the same department tend to communicate more, and
that the algorithm is capable of resolving separate centres to a good
degree of accuracy. The complicated branching structure resembles
self-similar systems in nature such as river networks or
diffusion-limited aggregates. {\bf b}, Same as before but without
showing the leaves. Branches are now coloured according to their
Horton-Strahler index (see text) {\bf c}, Binary tree showing the
result of applying the GN algorithm to a random graph with the same
size and connectivity than the e-mail network. The lack of community
structure is reflected in the absence of branches in the tree, which
contrasts with the intricate self-similar structure of {\bf a} and
{\bf b}. Again, colours correspond to Horton-Strahler indices.}
\label{arbre}
\end{figure}
The community binary tree for URV is shown in Fig.~\ref{arbre}. Each
colour in Fig.~\ref{arbre}a corresponds to one centre of the
university, that is to a faculty or college, or to management units
such as the office of the Rector of the university. Two properties of
the tree are worth noting. First, a clear branching structure emerges,
with branches essentially containing nodes of the same colour. This
shows that the identification of communities is successful, despite
the complexity of the network. Second, the branching structure is far
from simple. Indeed, each branch is formed, in general, by a system of
nested smaller subbranches that give rise to a complicated structure
that visually resembles some self-similar systems in nature such as
river networks \cite{rodriguez96} or diffusion-limited aggregates
\cite{halsey97}. For comparison, we also show the tree generated by
the GN algorithm from a random graph of the same size and average
connectivity as the e-mail network (Fig. \ref{arbre}c). In contrast to
the tree for the URV e-mail network, the branching structure is almost
trivial with most of the branches containing only 1 or 2 nodes. This
is the expected result for a network that do not have any sort of
community structure.

%\section{Self-similarity in the community structure}
\begin{figure}
\centerline{\includegraphics*[height=0.35\columnwidth]{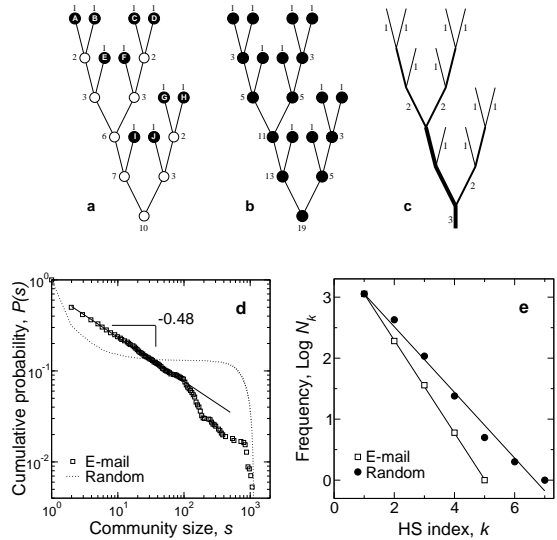}}
\vspace{0.5cm}
\centerline{\includegraphics*[width=0.40\columnwidth]{sizedis}\quad\includegraphics*[width=0.40\columnwidth]{hs}}
\caption{Self-similarity in the community structure. {\bf a},
Calculation of the community size distribution for a binary tree
generated by the community identification algorithm. Black nodes
represent the actual nodes of the original graph while white nodes are
just graphical representations of communities that arise as a result
of the splitting procedure. Nodes $A$ and $B$ belong to a community of
size 2, and together with $E$ form a community of size 3. Similarly,
$C$, $D$ and $F$ form another community of size 3. These two groups
together form a higher level community of size 6. Following up to
higher and higher levels, the community structure can be regarded as
the set of nested groups. The size, $s_i$, of a community $i$ is just
the summation of the sizes of its two offspring $j_1$ and $j_2$:
$s_i=s_{j_1} + s_{j_2}$. In this case there are three communities of
size 2, three communities of size 3, one community of size 6, one
community of size 7, and one community of size 10. Note that a single
node belongs to different communities at different levels. {\bf b},
Calculation of the drainage area distribution for a river network. The
drainage area of a given point is the number of nodes upstream of it
plus one. For a point $i$ with offspring $j_1$ and $j_2$, $s_i=s_{j_1}
+ s_{j_2} + 1$. {\bf c}, Calculation of the Horton-Strahler index. The
index of a branch changes when it meets a branch with higher index, or
when it meets a branch with the same value and both of them join
forming a branch with higher index. In this case, there are 10
branches with index 1, 3 branches with index 2, and 1 branch with
index 3. {\bf d}, The distribution of community sizes, $P(s)$, showing
a power law region with the exponent -0.48, followed by a sharp
decrease at $s\approx 100$ and a cutoff corresponding to the size of
the system at $s\approx 1000$. The distribution of community sizes in
a random network is shown with a dotted line for comparison. {\bf e},
The number of branches with HS index $i$, as a function of $i$. From
the definition of the branching ratio, it is straightforward to show
that, when topological self-similarity holds, $N_i=N_1 / B^{i-1}$. A
fitting of this function to the points obtained for the e-mail
community tree yields excellent agreement with $B=5.76$. A much worse
agreement is obtained for the community tree corresponding to the
random network, with $B_i$ fluctuating around 3.46.}
\label{area}
\end{figure}
Once the binary tree has been obtained, we look for a quantitative
characterisation of the community structure. First we consider the
cumulative community size distribution, $P(s)$, i.e. the probability
of a community having a size larger or equal to $s$. Fig.  \ref{area}a
shows how to compute this probability, and the resulting distribution
for the e-mail network is presented in Fig.  \ref{area}d. The
distribution is heavily skewed, following a power law behaviour
$P(s)\propto s^{-\alpha}$ with $\alpha=0.48$ between $s=2$ and
$s\approx 100$. Beyond this value, the distribution shows a sharp
decay and, at $s\approx 1000$, a cutoff that corresponds to the size
of the system. The power law of the community size distribution
suggests that there is no characteristic community size in the network
(up to $s\approx100$). To rule out the possibility that this behaviour
is due to the community identification algorithm, we also consider the
community size distribution for a random graph with the same size and
average connectivity as the e-mail network.

The characterisation of the community binary tree using the cumulative
size distribution has its analogy in the river network literature
\cite{rinaldo93,rodriguez96,maritan96}. The equivalent measure is the
distribution of drainage areas, that represents the amount of water
that is generated upstream of a given point (see Fig.
\ref{area}b). The similitude between the community size distribution
of the current e-mail network in Fig. \ref{area}d and the area
distribution of the Fella river network in Italy reported in Fig. 2 of
Ref.~\cite{maritan96} is striking. The exponent $\alpha=0.45$ for the
power law region of this river and the average exponent for several
rivers $\alpha_{river}=0.43\pm0.03$ respectively reported by
\cite{maritan96} and \cite{rinaldo93}, are very close to the current
$\alpha=0.48$. Moreover, the behaviour shown in Fig. \ref{area}d with
first a sharp decay and then a final cutoff is also shared by river
networks, which are known to evolve to a state where the total energy
expenditure is minimised \cite{rinaldo93,kramer92,sinclair96}. The
possibility that communities within organisations might also
spontaneously self-organise into a form in which some quantity is
optimised is very appealing and deserves further investigation.

To further understand this point, it is pertinent to ask the question:
does the similarity between community trees in organisations and river
networks arise just by chance or are there other emergent properties
shared by both?  To answer this question we consider a standard
measure for categorising binary trees: the Horton-Strahler (HS) index,
originally introduced for the study of river networks by Horton
\cite{horton45}, and later refined by Strahler
\cite{strahler52}. Consider the binary tree depicted in
Fig. \ref{area}c. The leaves of the tree are assigned a HS index
$i=1$. For any other branch that ramifies into two branches with HS
indices $i_1$ and $i_2$, the index is calculated as follows:
\begin{eqnarray*}
i=\left\{ \begin{array}
{l@{\quad \quad}l}
i_1 + 1       & \mbox{if}\quad i_1 = i_2,\\
\max(i_1,i_2) & \mbox{if}\quad i_1 \neq i_2.
\end{array} \right.
\label{HS}
\end{eqnarray*}
Note that the index of a branch changes when it meets a branch with
higher index, or when it meets a branch with the same value and both
of them join forming a branch with higher index. In terms of
communities, the interpretation of the HS index is the following. The
index of a community changes when it joins a community of the same
index. Consider, for instance, the lowest levels: individuals ($i=1$)
join to form a group (or team, with $i=2$), which in turn will join
other groups to form a {\it second level} group (or department,
$i=3$). Therefore, the index reflects the {\it level} of aggregation
of communities. The number of branches $N_i$ with index $i$ can be
determined once the HS index of each branch is known . The bifurcation
ratios $B_i$ are then defined by $B_i=N_i / N_{i+1}$ (by definition
$B_i\ge{2}$).

When $B_i\approx B$ for all $i$, the structure is said to be
topologically self-similar, because the overall tree can be viewed as
being comprised of $B$ sub-trees, which in turn are comprised of $B$
smaller sub-trees with similar structures and so forth for all
scales. River networks are found to be topologically self-similar with
$3<B<5$ \cite{halsey97}. We find that the community tree as generated
by the process described above is topologically self-similar with
$B_i\approx B=5.76$ (see Fig. \ref{area}e). The same analysis for
the communities in a random graph shows that topological
self-similarity does not hold, since the values $B_i$ are not
constant; they fluctuate around a smaller 3.46 value.

%\section{Conclusions}
The methods presented here open interesting doors regarding the
possibility of mapping the informal network of large organisations in
a non-intrusive, objective, and quantitative way. Moreover, the
emergence of scaling and self-similarity in the community structure,
as well as the similarity with river networks, raises important
questions about the mechanisms underlying the interactions between
individuals within an organisation. Self-similarity is a fingerprint
of the replication of the structure at different levels of
organisation, and could be the result of the trade-off between the
need for cooperation and the physical constrains to establish
connections at any organisational level. At the same time, the
similitude with river networks suggests that a common principle of
optimisation---of flow of information in organisations or of flow of
water in rivers---could be the underlying {\it driving force} in the
formation and evolution of informal networks in organisations.

\acknowledgments We thank L. Amaral, M. Buchanann, X. Guardiola, and
Julio Ottino for helpful comments and suggestions. We also thank
J.~Tomas, O.~Lorenzo, C.~Llorach, J.~Clavero and F.~Salvador for
collecting the e-mail data. This work has been supported by DGES of
the Spanish Government and EC-Fet Open project COSIN. R.G. and
L.D. also acknowledge financial support from the Generalitat de
Catalunya.

%%%%%%%%%%%%%%%%%%%%%%%%%%%%%%%%%%%%%%%%%%%%%%%%%%%%%%%%%%%%%%%%%%%%%%%
%%%%%%%%%%%%%%%%%%%%%%%%%%%%%%%%%%%%%%%%%%%%%%%%%%%%%%%%%%%%%%%%%%%%%%%
%%%%%%%%%%%%%%%%%%%%%%%%%%%%%%%%%%%%%%%%%%%%%%%%%%%%%%%%%%%%%%%%%%%%%%%
%%%%%%%%%%%%%%%%%%%%%%%%%%%%%%%%%%%%%%%%% REFERENCES

%\bibliography{/home/rguimera/Recerca/Textos/bibliografia}

\end{document}